\documentclass[aps,prb,preprint,superscriptaddress]{revtex4-1}

\usepackage{graphicx} 
\usepackage{graphics} 
\usepackage{amsfonts} 
\usepackage{caption}
\usepackage{subcaption}
\usepackage{amsmath}
\usepackage{amssymb} %
\usepackage{color} 
\usepackage{epstopdf}
\usepackage{hyperref}
\usepackage{lineno}

\hypersetup{
	colorlinks = true,
	allcolors = {blue},
	allbordercolors = {1 1 1}
}

 \newcommand{\ket}[1]{\mbox{$|#1\protect\rangle$}}
 \newcommand{\bra}[1]{\mbox{$\protect\langle#1|$}}
 
 \newcommand{\expect}[1]{\mbox{$\protect\langle #1 \protect\rangle$}}

\begin{document}

\title{How the Result of Counting One Photon Can Turn Out to Be a Value of 8}

\author{Matin \surname{Hallaji}}
\author{Amir \surname{Feizpour}}
\author{Greg \surname{Dmochowski}}
\author{Josiah \surname{Sinclair}}
 \affiliation{Centre for Quantum Information and Quantum Control and Institute for Optical Sciences, Department
of Physics, University of Toronto, 60 St. George Street, Toronto, Ontario,
Canada M5S 1A7}
\author{Aephraim M. \surname{Steinberg}}
 \affiliation{Centre for Quantum Information and Quantum Control and Institute for Optical Sciences, Department
of Physics, University of Toronto, 60 St. George Street, Toronto, Ontario,
Canada M5S 1A7}
\affiliation{Canadian Institute For Advanced Research, 180 Dundas St. W., Toronto Ontario, CANADA M5G 1Z8}

\maketitle

\textbf{In 1988, Aharonov, Albert, and Vaidman introduced a new paradigm of quantum measurement in a paper which had the unwieldy but provocative title ``How the result of a measurement of a component of the spin of a spin-$1/2$ particle can turn out to be $100$."\cite{AAV}  This paradigm, so-called ``weak measurement," has since been the subject of widespread theoretical and experimental attention, both for the perspective it offers on quantum reality and for possible applications to precision measurement.  Yet almost all of the weak-measurement experiments carried out so far could be alternatively understood in terms of the classical (electromagnetic wave) theory of optics.  Here we present a truly quantum version, the first in which a measurement apparatus deterministically entangles two distinct optical beams, enabling us to experimentally ask a question directly analogous to that of the original proposal:  
``In a two-arm interferometer containing one photon in total, can the result of a measurement of the photon number in one arm turn out to be greater than $1$?''
Specifically, we show that a single photon, when properly post-selected, can have an effect equal to that of eight photons: that is, in a system where a single photon has been calibrated to write a nonlinear phase shift of $\phi_0$ on a ``probe beam," we measure phase shifts as large as $8\phi_0$ for appropriately post-selected single photons.  This is the first deterministic weak-value experiment in optics which defies classical explanation, and constitutes a realization of our proposal for ``weak-value amplification" (WVA) of the small optical nonlinearity at the single-photon level\cite{amir}.  It opens up a new regime for the study of entanglement of optical beams, as well as further investigations of the power of WVA for the measurement of small quantities.}

Physical measurement of a property of a system generally proceeds by coupling the system to a probe in such a way that the change of the state of the probe depends on the value of this property; for example, a galvanometer is constructed so that its needle deflects by an amount proportional to the potential difference across the system being studied. 
A subsequent observation of the final state of the probe provides information (often incomplete) about the value of the observable. 
This information, however, is gained at the price of disturbing the system through the interaction; in quantum mechanics, as is well known, there is a strict trade-off between the minimum disturbance and the amount of information which can be gained\cite{PhysRevLett.109.100404,PhysRevA.88.022110}.
In weak measurement, the disturbance to the system is reduced, at the cost of a similar reduction in the amount of information provided by the measurement. 
This minimal disturbance makes it reasonable to consider conditioning the read-out of the probe on finding the system in a particular final state after the interaction (post-selection.)
In this case, the pointer shift, averaged over many measurement repetitions, has been shown\cite{AAV} to have a magnitude which would correspond to what is termed the ``weak value'' of the observable: $\langle f | \hat{A} | i \rangle / \langle f | i \rangle$,  where $\hat{A}$ is the observable, and $| i \rangle$ and $| f \rangle$ are the pre- and post-selected states of the system, respectively. 
Evidently, the weak value depends equally on both pre-selected (initial) and post-selected (final) states.
This feature of weak measurement makes it a powerful tool for exploring fundamental questions in quantum mechanics \cite{box,traj,heisenberg,erhart2012experimental,lundeen,PhysRevLett.85.3149,hardy,1367-2630-11-3-033011,PhysRevLett.111.240402,PhysRevLett.116.180401,prob_interact,denkmayr2014observation,Murch2016}, and specifically the properties of post-selected subensembles ranging from particles transmitted through tunnel barriers to measurement-based quantum-computing systems \cite{tunneling,KLM,cluster}.
Strangely, the weak value is not constrained to be within the eigenvalue spectrum of the observable $\hat{A}$, and is not even in general a real number. 
In particular, as the overlap between the initial and final states becomes very small, $\langle f | i \rangle \rightarrow 0$, the weak value can become (almost) arbitrarily large (as long as the post-selection success is dominated by the overlap of pre- and post-selected states and not the measurement back-action; see the supplementary materials.)
This has led to the idea of using ``weak-value amplification'' (WVA) to improve the detection or measurement of small effects \cite{kwiat,jordan_howell,Pfeifer:11,PhysRevLett.111.023604,Jayaswal:14,PhysRevLett.112.200401,PhysRevA.89.012126}.
Interest in this application of weak measurement has grown in the past few years alongside an ongoing debate on the usefulness of WVA\cite{yamamoto,FandC_sub,lessismore,FandC_lim,howell_exp_improv}.
Even the quantum mechanical nature of WVA has been challenged\cite{FandC_coin} and attempts have been made to describe the effect classically based on measurement disturbance.

Anomalous weak values observed to date\cite{kwiat,jordan_howell} have typically utilized two different degrees of freedom (such as polarization and propagation direction) of a photon as the ``system" and the ``probe," obviating the need for any inter-photon interaction; the effects can thus be explained perfectly in terms of linear optics, without resorting to quantum theory. 
There have been two exceptions.  
In one, a probabilistic quantum logic gate was implemented, so that although there was no deterministic entanglement of system and probe, an additional post-selection step projected the system onto an entangled state some fraction of the time\cite{prob_interact,LG_ineq_photons}.
In the other, deterministic WVA was implemented in a transmon qubit system\cite{deter_interact}.
Here, we present the first observation of WVA via deterministic entanglement of two distinct optical systems, ``amplifying'' the number of photons in a signal beam by measuring the nonlinear phase-shift it writes on a separate probe beam. 

\begin{figure}[h]
\centering
\includegraphics[width = 0.7\columnwidth]{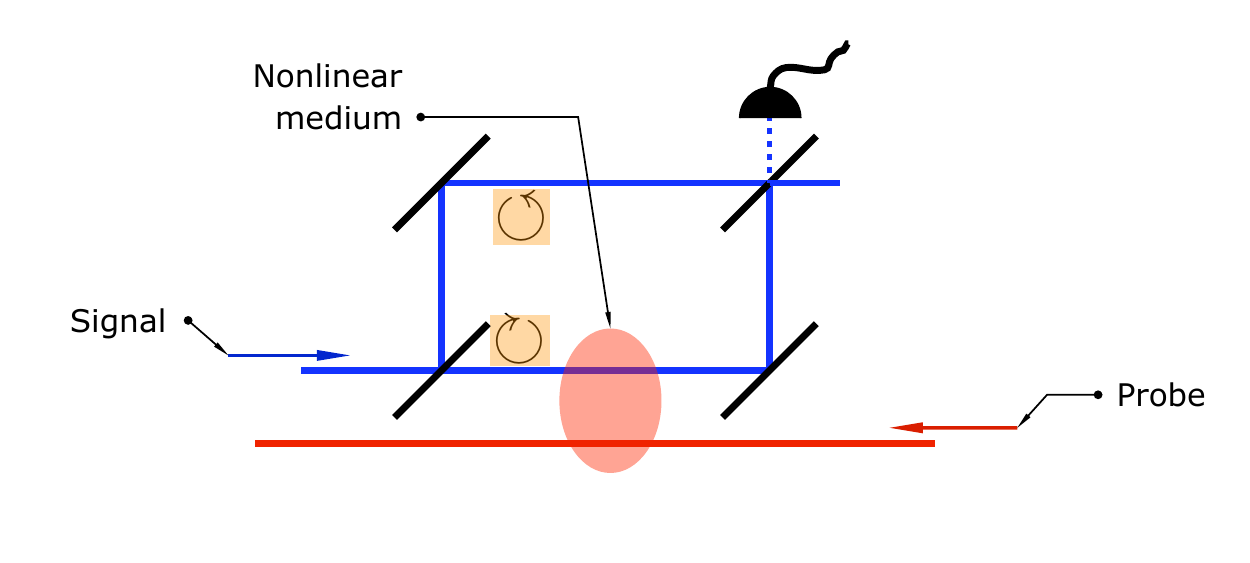}
\caption{
  \label{fig_setup}}
\end{figure}

In 2011, we proposed that WVA could be used to amplify the (weak) nonlinear effect of photons in a signal pulse on a probe beam.
The scheme begins by splitting the signal beam into two paths (see Fig. \ref{fig_setup}), which are later interferometrically recombined such that there is strong constructive interference at one port, and very few photons exiting the other, ``nearly-dark,'' port.  By post-selecting on cases where a photon exits this nearly-dark port, one can ``amplify'' the weak value of the photon number in one arm of the interferometer. 
The physical content of this statement is that if an optical probe beam interacts, through a Kerr-type nonlinear medium, with the light in that arm of the interferometer, it will experience a phase shift proportional to the weak value of the photon number, which may be much larger than the nonlinear phase shift expected for a single signal photon -- even if there was only ever a single signal photon present in the entire interferometer.

%

\begin{figure}[h]
\centering
\includegraphics[width = \columnwidth]{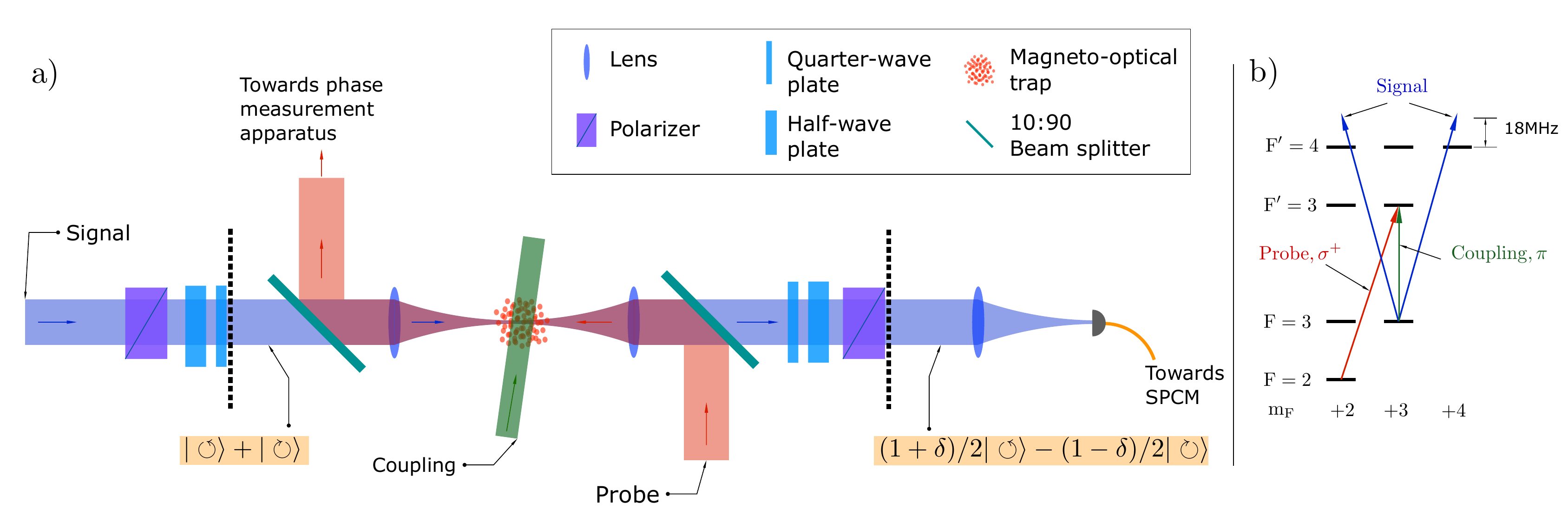}
\caption{ 
  \label{xps_pol}}
\end{figure}

In our experiment, the nonlinear interaction between the signal and probe beams is mediated by a sample of laser-cooled $^{85}$Rb atoms in a magneto-optical trap (MOT). 
A coupling beam is used to set up Electromagnetically Induced Transparency (EIT) for the probe beam.
The frequencies of the probe and coupling beams are set so that each of them is individually on resonance with its corresponding atomic level; see Fig.\ref{xps_pol}b. 
The EIT resonantly enhances the nonlinear interaction between the signal and the probe beams, while simultaneously minimizing the probe absorption. 
As a signal pulse passes through the medium, it alters the optical properties (the index of refraction) of the medium as seen by the probe beam. 
As a result of this change in the index of refraction, the probe picks up a phase shift relative to what the phase would have been in the absence of the signal pulse. 
This cross phase shift (XPS) depends linearly on the number of photons (intensity) of the signal pulse\cite{PSXPS}.
This interaction thus constitutes a measurement of signal photon number, with the phase of the probe beam acting as a ``pointer.''  
The typical phase shift per photon is on the order of $10^{-5}$ rad\cite{PSXPS}, much smaller than the quantum uncertainty in our probe phase, meaning that a single measurement cannot provide enough information to determine the photon number to better accuracy than its initial uncertainty, which is the sense in which this measurement is ``weak.''

The geometry and polarizations of the coupling and probe beams (Fig.\ref{xps_pol}) are chosen such that the strength of this nonlinear interaction also depends on the polarization of the signal beam, with the interaction being strongest for right-handed circular ($\sigma ^{+}$) polarization and weakest for left-handed circular ($\sigma ^{-}$) polarization of the signal; see Supplementary Material.
As mentioned before, an interferometer is necessary in order to implement WVA of signal photon number.
We chose this configuration so as to use the polarization dependence of XPS to build an interferometer for the signal beam, in which the two ``paths'' of the interferometer are in fact two polarizations, to one of which the probe beam is much strongly coupled than to the other (see Fig.\ref{xps_pol})
The incident signal photons are linearly polarized, i.e., in an equal superposition of $\sigma ^{+}$ and $\sigma ^{-}$.
After they interact with the probe beam, we use a waveplate and a polarizer to transmit photons with a nearly-orthogonal polarization (corresponding to the nearly dark interferometer port, see Fig.\ref{fig_setup}), which subsequently impinge on a single-photon counting module (SPCM).  When this detector fires, constituting a successful post-selection, a large weak value of photon number (and hence large nonlinear phase shift) is expected; for each signal pulse, we record whether or not the detector fires, in order to measure the average nonlinear phase shift written on the probe separately for the cases of successful post-selection (which we term ``click'') and the cases of failed post-selection (``no-click'').



Although our original proposal concerned a signal pulse that contained  a single-photon Fock state\cite{amir}, a similar effect can be observed even if one uses a coherent state, with an average photon number larger than one. 
In that proposal we showed that the weak value of photon number in one arm of the interferometer, when the photon is post-selected to be in the nearly-dark port, is given by $\langle n \rangle _{wk} = 1/2 + 1/2\delta$.
The parameter $\delta$ is defined as the overlap between the initial and final state of the photon, ranging from 0 (completely dark port) to 1 (bright port).
However, here $\delta$ is assumed to be very small, which corresponds to a nearly-dark port.
In another work\cite{PSXPS}, we showed that, in the limit of low detection efficiency, the (mean) inferred photon number in a coherent state containing $\bar{n}$ photons, when conditioned upon a successful photon detection, increases by one photon: the inferred photon number after a successful post-selection, $n_{\mathrm{click}}$, turns out to be 1 more than the inferred photon number in the absence of a click, $n_{\mathrm{no-click}}$.
Now, if one combines these ideas, sending a coherent state in the interferometer instead of a single photon Fock state and conditioning the photon number measurement of one arm on detection of a photon in the nearly-dark port, the added photon due to the photon detection will undergo weak value amplification.
Using B and D to denote bright and nearly-dark ports, one can approximate the initial signal coherent state $|i\rangle \approx |\alpha\rangle_B \, |\alpha \delta \rangle_D$ as $|\alpha\rangle_B \, \left(|0\rangle + \alpha \delta|1\rangle\right)_D$, and the final state post-selected when the detector fires as $ |\alpha\rangle_B \, |1\rangle_D$ (occurring with probability $|\alpha|^2 |\delta|^2$, multiplied by the experimental collection and detector efficiency.)
One can then calculate the weak value of the number of photons in each arm of the interferometer ,$\langle n_{\pm}\rangle _{wk}$, with ``plus'' and ``minus'' signs denoting the $\sigma^+$ and $\sigma^-$ polarizations, by 
\begin{equation}
\langle n_{\pm}\rangle _{wk} = \frac{\langle f | \hat{n}_{\pm} | i \rangle}{\langle f | i \rangle}.
\end{equation}
The number operators $\hat{n}_{\pm}$ can be written $a^\dagger_\pm a_\pm$, where the field operators $a_\pm \approx \left (a_B \pm a_D\right) / \sqrt{2}$.
The numerator can thus be written 
\begin{equation}
\begin{split}
\frac{1}{2}\;\bra{\alpha}_B\bra{1}_D [a^\dagger_B \pm a^\dagger_D] & \cdot  [a_B \pm a_D] \left\{\ket{0}_D+\alpha \delta \ket{1}_D\right\}\ket{\alpha}_B \\
= \frac{1}{2} \; \big[ \bra{\alpha}_B\bra{1}_D\alpha^* \pm \bra{\alpha}_B\bra{0}_D \big] & \cdot \big[ \left(\alpha \pm \alpha \delta\right) \ket{\alpha}_B\ket{0}_D + \alpha^2 \delta \ket{\alpha}_B\ket{1}_D \big] \;.
\end{split}
\end{equation}
As the denominator simply evaluates to $\alpha \delta$, we find 
\begin{equation}
\label{WVA_equation}
\expect{n_\pm}_{\rm wk} = \frac{1}{2}\; \frac{\alpha |\alpha|^2 \delta +\alpha  \delta  \pm\alpha }{\alpha \delta} = \frac{1}{2}\left(|\alpha|^2 + 1 \pm 1/\delta\right).
\end{equation}
(See the supplementary materials for a more detailed derivation of Eq.\ref{WVA_equation}.)
It is instructive to note that the total number of photons inside the interferometer is $\langle n_ {+}\rangle _{wk} + \langle n_{-}\rangle _{wk} = \bar{n} + 1$, where $\bar{n}$ is the average number of photons sent into the interferometer, and $\bar{n}+1$ is the revised estimate of the mean photon number based on detection of one ``additional'' photon in the nearly-dark port.
What equation \ref{WVA_equation} demonstrates is that this added photon (``the'' photon which causes the SPCM in the nearly-dark port to fire, so to speak) undergoes WVA, giving rise to the term $1/2\delta$; this occurs even in the presence of a ``background'' of $\bar{n}$ non-post-selected photons, which experience no amplification effect.
In this context, when the probe beam interacts with the one arm of the interferometer, the nonlinear phase shift written on it will be proportional to the weak value of photon number in that arm.
Due to the spatial overlap between the probe beam and the two arms, which are the two circular polarizations of the signal beam, the total XPS written on the probe is $\phi = \langle n_ {+}\rangle _{wk} \phi_+ + \langle n_ {-}\rangle _{wk}\phi_-$, where $\phi_{\pm}$ are the per-photon phase shifts for the corresponding signal polarizations.
Upon detection of a photon in the nearly-dark port, a click event, these are to be replaced by the weak values from equation \ref{WVA_equation}, leading to $\phi_{\rm click} = (\bar{n}+1) \frac{\phi_+ + \phi_-}{2}+\left(\frac{1}{\delta}\right) \frac{\phi_+ - \phi_-}{2 }$.  
The amplification shows up in the second term, which is proportional to the difference between the XPS for the two polarizations.
In cases when the detector fails to detect a photon, no-click events, there is no ``added photon'' and the weak value of photon number in each arm is simply $\bar{n}/2$, corresponding to an XPS of $\phi_{\rm no-click} = \bar{n} \frac{\phi_+ + \phi_-}{2}$.

\begin{figure}[h]
\centering
\includegraphics[width = \columnwidth]{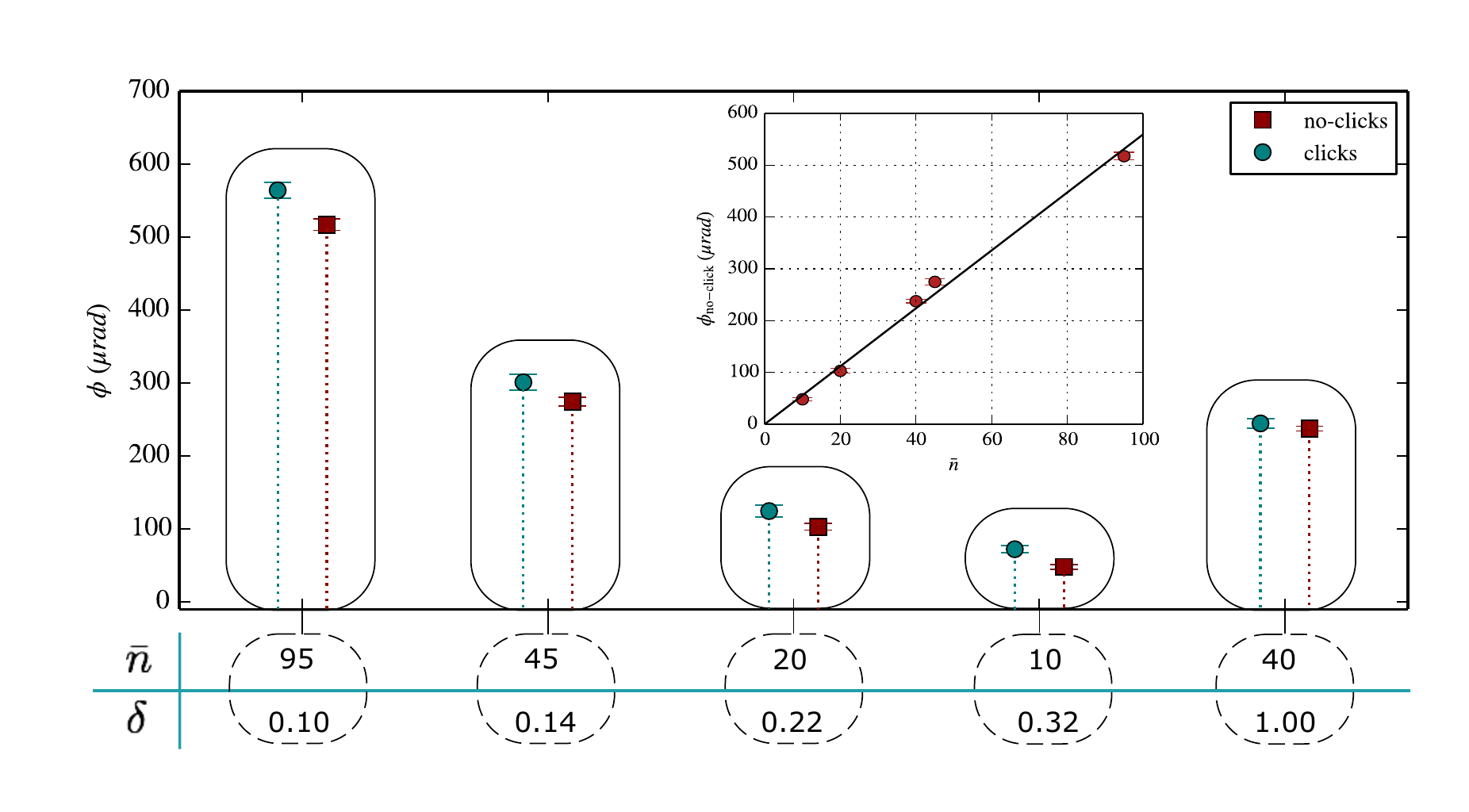}
\caption{
  \label{yes_no}}
\end{figure}

Figure \ref{yes_no} shows the measured XPS for the click and no-click events versus a range of values of post-selection parameter $\delta$; 
for each $\delta$, the mean input photon number $\bar{n}$ and the overall detection efficiency $\eta$ are adjusted so as to keep the probability of photon detection ($P_{click} = \eta \delta^2 \bar{n}$) low, which is the necessary condition for a photon detection to add one photon to the inferred photon number\cite{PSXPS} (for more technical details see the supplementary materials.)
The measured XPS for the click events is manifestly always larger than the measured XPS for the no-click cases.
It is worth noting that had we not utilized post-selection to separate the click from no-click events, the expected XPS of the probe would have been $\phi = \bar{n} \frac{\phi_+ + \phi_-}{2}$, which is the same as $\phi_{\rm no-click}$ when the overall detection efficiency is low.
Therefore, we can define $\phi_0 = \frac{\phi_+ + \phi_-}{2}$ as the expected per-photon phase shift for this experiment.
The inset of Fig. \ref{yes_no} plots $\phi_{\rm no-click}$ versus the input photon number.
From a linear fit to this data, the per-photon phase shift  $\phi_0$ is measured to be $5.59 \pm 0.02 \mu$rad.

\begin{figure}[h]
\centering
\includegraphics[width = \columnwidth]{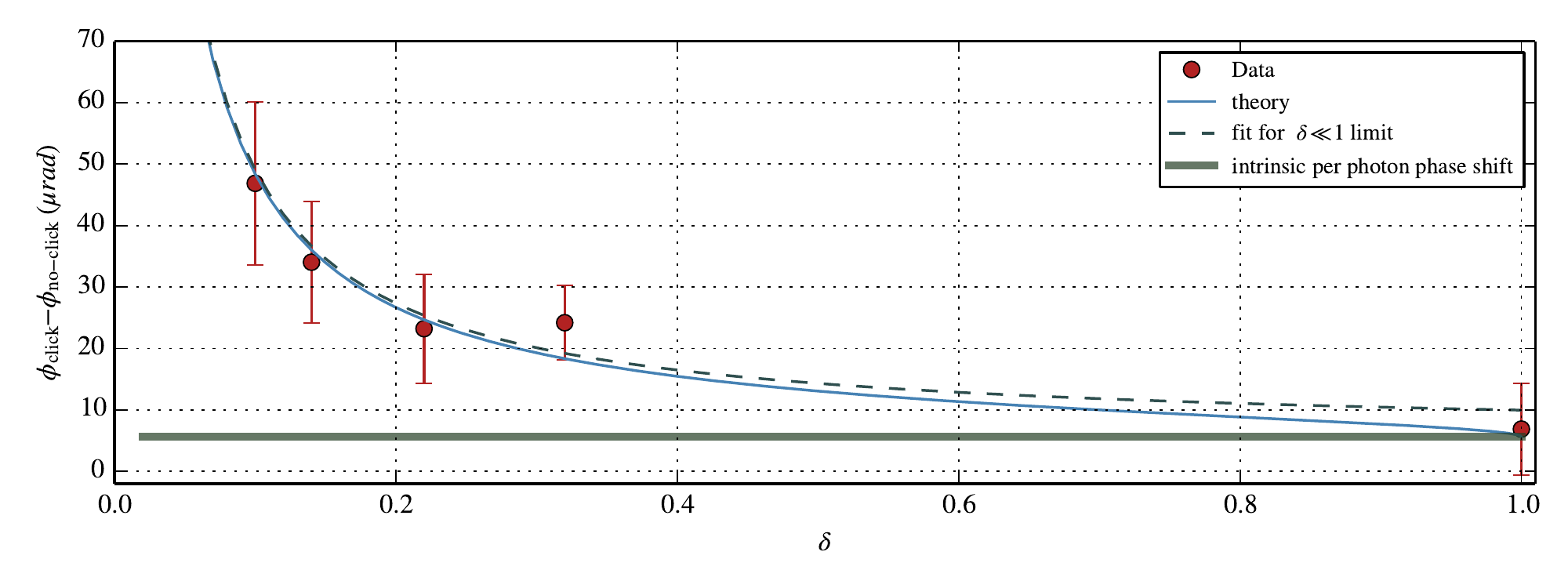}
\caption{ 
  \label{xps_del}}
\end{figure}

To directly observe the WVA of the added photon due to photon detection, we plot $\phi_{click} - \phi_{no-click}$ versus the post-selection parameter $\delta$ in Fig. \ref{xps_del}. 
It is easy to see that this quantity, which we term ``the differential phase shift", is independent of $\bar{n}$, and only contains the XPS of the added photon and its amplified effect: $\phi_{click} - \phi_{no-click} = \frac{\phi_+ + \phi_-}{2} + \left(\frac{1}{\delta}\right) \frac{\phi_+ - \phi_-}{2 }$.
The plot clearly shows that as $\delta$ becomes smaller, the effect of the post-selected single photon becomes larger.
For $\delta = 0.1$, the smallest post-selection parameter used in our experiment, we measure a differential phase shift of $47.0 \pm 13.5 \mu$rad, which is 8.4$\pm$2.4 times larger than the per-photon phase shift $\phi_0$.
Hence, a single post-selected photon can act like 8 photons. 
In an ideal case, where only the photon number in one arm of the interferometer is measured, $\delta = 0.1$ corresponds to an amplification factor of 10.  
For the points with $\delta=0.14$ and $\delta=1$, we used nearly the same mean photon number ($45$ and $40$, respectively), so that the results would be directly comparable even without knowledge of the independence of the differential phase shift on $\bar{n}$.  
For the $\delta=1$ case, we observed $6.7 \pm 7.5 \mu$rad, consistent with the unamplified value of $5.59 \pm 0.02 \mu$rad expected in the absence of WVA, while for the $\delta=0.14$ case we found $34 \pm 10 \mu $rad, 3 standard deviations above $\phi_0$.
From a fit to the data in Fig. \ref{xps_del} we estimate $\phi_+ - \phi_-$ to be 8.7$\pm$0.6$\mu$rad.

Given the extensive discussion in recent years over the possible merits of WVA for making sensitive measurements of small parameters, it is interesting to contrast the present experiment with an earlier one, in which we measured the nonlinear phase shift due to post-selected single-photons, but without any weak-value amplification\cite{PSXPS}.
In our previous experiment, a total of approximately 1 billion trials (300 million events with post-selected photons, and 700 million without) were used to measure the XPS due to  $\sigma^+$-polarized photons. 
By looking at the difference between the XPS measured for ``click'' and ``no-click'' events, we measured peak XPS $\phi _+$ of $18 \pm 4 \mu$rad.
In this experiment, where we use the WVA technique, we used a total of around 830 million trials (200 million successful post-selections) to extract an average XPS $\phi_+$ of 10.0$\pm$0.6$\mu$rad (for more information regarding the reported average XPS see the Probe phase measurement section in the supplementary material).
Note that this number it agrees well with our classical calibration of the peak XPS of 13.0$\pm$1.5$\mu$rad\cite{PSXPS}.
It is evident that the WVA technique yielded a better signal-to-noise ratio (SNR.)
This may seem surprising at first, given that under statistical-noise conditions, WVA is known to have the same SNR as a brute-force measurement\cite{amir}; this is because the amplification of the signal only comes at the price of a post-selection which reduces the size of the data set just enough to cancel out any advantage one might have hoped for. 
In our case, however, while the differential phase shift grows as $1/\delta$, the size of the post-selected data set is determined by $\delta^2 \bar{n}$.
Hence, we were able to maintain a substantial set of post-selections even for small $\delta$ simply by adjusting $\bar{n}$ accordingly.
This allowed us to amplify the per-event signal while keeping the number of events high, thereby achieving a precise measurement of the differential phase shift.  
Furthermore, in the no-click cases, the large photon number provided an excellent determination of $\phi_0$.
Therefore, we were able to estimate the value of $\phi _+$ more precisely with fewer trials.
It is essential to note that the WVA technique in this experiment is advantageous only if one considers the number of trials (and not the number of photons) as the measurement resource.  

This work unambiguously demonstrates the power of weak measurement to amplify the effect of a single post-selected photon on a probe beam, the first weak measurement carried out with a deterministic optical interaction allowing no classical explanation.  
Future work will address the ongoing discussions about the advantages of WVA over traditional strong measurement.  
To tackle this question, future iterations of the experiment can be carried out with artificial noise added, conforming to a variety of noise models. 
It is expected that in cases where the noise has long time correlations, WVA will prove superior to standard measurement for the estimation of small parameters such as the intrinsic per-photon phase shift or $\chi ^{(3)}$ of a sample.

\newpage
\textbf{Figure captions}

\emph{Figure \ref{fig_setup}}.
Conceptual schematic of the interferometer. The signal beam is split into two paths, labeled $\circlearrowleft$ and $\circlearrowright$(to make a connection with the actual, polarisation-based, interferometer shown in Fig. \ref{xps_pol}).
A probe beam measures the number of photons in one path through a nonlinear interaction. The interferometer is made slightly imbalanced so that there is a small chance for a signal photon to be detected in the nearly-dark port.

\emph{Figure \ref{xps_pol}}.
a) The experimental setup and our implementation of the polarization interferometer. 
Counter-propagating probe and signal beams are focused to a waist of $13\pm1\mu$m inside a cloud of laser cooled $^{85}$Rb atoms confined in a magneto-optical trap (MOT).
90\%-reflectivity beam-splitters are used to overlap the probe beam with and separate it from the signal beam.
The phase of the probe is then measured using frequency domain interferometry.
A collimated coupling beam, propagating perpendicularly to the probe and signal beams, generates EIT for the probe.
The polarization of the signal beam is set to be linear before the interaction, i.e. an equal superposition of right-handed and left-handed circular polarizations $|\circlearrowleft \rangle + |\circlearrowright \rangle $.
After the interaction the polarization of the signal beam is post-selected to be in the final state $\frac{(1+\delta)}{\sqrt{2}}|\circlearrowleft \rangle - \frac{(1-\delta)}{\sqrt{2}}|\circlearrowright \rangle $, where $\delta$ denotes the overlap between the two polarization states.
A single photon counting module (SPCM) is used to detect the presence of a photon in the dark port.  
The vertical dashed lines mark the beginning and the end of the interferometer.
b) Atomic level structure used for the experiment.
Only the levels with the highest interaction strengths are shown.
See the supplementary material for the full level structure.

\emph{Figure \ref{yes_no}}.
Measured XPS for click events and no-click events.
Red squares and green circles are the measured phase shift conditioned on the SPCM not firing (no-click) and firing (click) respectively.
The horizontal axis shows the mean signal photon number $\bar{n}$ and the post-selection parameter $\delta$ used for each case.
The inset plots the measured phase shift for the no-click cases versus the average photon number in the signal beam $\bar{n}$.
The linear fit reveals a per-photon XPS of $\phi_0 = 5.59 \pm 0.02 \mu$rad.

\emph{Figure \ref{xps_del}}.
Differential phase shift versus post-selection parameter.
Red circles are the phase shift difference between click and no-click cases. 
The black dashed line is a fit to $\frac{\phi_+ + \phi_-}{2} + \frac{\phi_+ - \phi_-}{2 \delta}$, where the measured value of $5.59 \mu$rad (shown as solid green line) is used for $\frac{\phi_+ + \phi_-}{2}$ and $\phi_+ - \phi_-$ is left as a fitting parameter. 
The best fit corresponds to a value of $8.7 \pm 0.6 \mu$rad for $\phi_+ - \phi_-$.
The fit is only valid for $\delta \ll 1$.
The blue solid line is the full theoretical calculation assuming the numbers mentioned above.

\newpage
\textbf{Methods}

\emph{Atom preparation.}
A cloud of $^{85}$Rb atoms is prepared in a magneto-optical trap (MOT).
Three pairs of beams are used for cooling while a magnetic gradient of approximately 20G/cm (along the quadrupole axis) provides confinement in space.
Each beam contains a trapping beam, tuned 20 MHz to the red of the $F=3\rightarrow F'=4$ transition, and a repumper, which is tuned close to the $F=2\rightarrow F'=3$ resonance.
Each measurement cycle lasts for 52ms. 
For the first 50ms, the MOT beams and magnetic field gradient cool and prepare the cloud.
They are then turned off and the atoms are probed for 1.5ms with a 500$\mu$s gap between cooling and probing to ensure the absence of residual magnetic field gradients.

\emph{Probe phase measurement.}
To measure the phase and amplitude of the probe beam, we use frequency-domain interferometry.
The probe beam is comprised of two frequency components that co-propagate through the interaction region.
One frequency component is tuned to resonance with $F=2 \rightarrow F'=3$.
A second frequency component is prepared by frequency-modulating an AOM at 100MHz, and is used as a probe reference.  
The beating of these two components is detected on a fast avalanche photodiode.
Any change in the phase and/or amplitude of the probe beam results in a change in phase and/or amplitude of this 100MHz beating signal.
We use an $IQ$-demodulator to then read off the phase ($\arctan(\frac{I}{Q})$) and amplitude ($\sqrt{I^2+Q^2}$) of the beat signal.
The resulting phase and amplitude are digitized with a sampling period of $\frac{200}{3}$ns (sampling frequency of 15MHz).
Each measurement cycle contains 22501 samples (corresponding to 1.5ms of phase measurement), which we call a trace.
The first 4500 samples (300$\mu$s) in each trace are ignored in order to avoid contamination from the residual probe phase dynamics.
The last 180 samples are used to measure the final OD of the atomic cloud (by turning the coupling beam off, hence eliminating the EIT, and measuring the probe absorption).
The remaining 1.188ms is divided into 495 shots, 2.4$\mu$s (36 samples) each.
Each shot contains a nonlinear phase shift due to the interaction between the probe and the signal pulse.
The nonlinear phase shift has a FWHM width of around 500ns\cite{PhysRevLett.116.173002}.
We average the value of this phase shift for 7 samples (466.6ns) in each shot. 
To eliminate slow drifts in our phase measurement, the phase of the probe is averaged for 3 samples (200ns) with a 266ns gap before and after the 7 samples.
This value is then subtracted from the average of the probe phase in the interval when nonlinear phase shift is expected.
The resulting number represents the average phase shift due to the signal pulses and is reported as the measured XPS.
It is worth noting that the reported values for XPS in this report underestimate the peak phase shift because of the averaging and background subtraction.
The probe beam contains about 2000 photons, which corresponds to a 11mrad quantum limit on the phase uncertainty (shot-noise.)
We measure a single-shot phase uncertainty of around 100mrad. 
In order to measure the phase shift down to few $\mu$rad precision, we repeat the measurement about half a billion times, with around 200 shots in each 1.5ms measurement window.

\emph{Probe and coupling fields.}
Electromagnetically induced transparency (EIT) is a coherent effect in which destructive interference prevents the two laser beams from being absorbed by the atoms.
In order to generate EIT, the probe and coupling lasers should be phase-locked.
To generate the probe beam, some power is first extracted from a master laser beam that is locked $\approx$30MHz red of the $F=2 \rightarrow F'=3$ transition.
By using an acousto-optic modulator (AOM) that is driven at +130MHz (double-passing at +65MHz) the off-resonance component of the probe beam is generated.
This off-resonance beam is then sent through another AOM at -100MHz to generate the on-resonance probe component.
The two beams are then combined on a beam splitter and then sent towards the interaction region.
We use an electro-optic modulator (EOM), which is driven at around 3GHz, to frequency modulate the remaining portion of the master laser; this writes frequency sidebands on the laser. 
The frequency modulated beam is then used to seed an injection-locked diode laser and we lock this diode laser to the first lower sideband.
As a result, the second diode laser, the coupling laser, is phase-locked to the master laser.
The frequency of the coupling beam is set to be on resonance with the $F=3 \rightarrow F'=3$ transition.
We use another AOM (single pass driven at +103 MHz) to switch the coupling beam on and off.
The intensities of the probe and coupling beams are chosen so that the resulting EIT width is 2MHz.
The polarizations of the probe and coupling beams are set to be $\sigma^+$ and $\pi$ respectively.

\emph{Signal pulses.}
A portion of the injection-locked diode laser, mentioned above, is sent though two AOMs. 
These AOMs are used to set the frequency of the pulses to be around +18MHz from the $F=3\rightarrow F'=4$ transition.
One of the AOMs is also used to amplitude-modulate the signal beam in order to create 40ns pulses. 
An ND (neutral density) filter is used to attenuate the signal pulses, preparing pulses with low average photon numbers.
A polarizer followed by a half-wave plate and a quarter-wave plate is used to set the polarization of the signal pulse before its interaction with the probe. 
For all WVA measurements reported in this Report, the signal pulse is initially linearly polarized.

\emph{Signal post-selection.}
After its interaction with the probe beam, the signal beam is sent through a half-wave plate, a quarter-wave plate and a Glan-Thompson polarizer.
With this combination, the polarization of the signal beam is fully characterized with and without the atoms (by doing tomography).
Once the initial polarization of the signal beam is well understood, and any polarization rotations due to the presence of atoms are corrected for, the half- and quarter-wave plates are set so that the signal polarization in the output of the Glan-Thompson polarizer is projected onto a polarization almost orthogonal to the initial polarization with real overlap with the initial polarization.
The signal pulses are then collected in a multi-mode fiber and are detected on an SPCM.
Upon detecting a photon, the SPCM sends a signal which triggers our tagging module to expose the probe detector to a 100ns flash of light.
This pulse shows itself as a spike in the probe amplitude and phase. 
This spike tags the corresponding shot as a successful post-selection.
A time delay between when the tag appears in a shot and when the XPS is expected to happen is introduced and carefully adjusted to avoid any incursion of the tags to the measured XPS in that shot.
The tags, however, affect the XPS in the next shot and, therefore, the shot after each tag is discarded.
We then use these tags, and the absence thereof, to group the shots into successful (click) and unsuccessful (no-click) post-selection bins.

\emph{Background photons.}
Any residual photon that hits the SPCM results in a photon detection which will falsely be counted as a successful post-selection.
In order to reduce the chances of getting a false positive, we time-gate the SCPM in the 40ns windows where we expect the signal pulses to arrive.
With this gating, 6$\%$ of the measured shots are still falsely tagged as `click'.
These background detections deteriorate the desired effect.
Therefore, to be less sensitive to these detections, we attempt to operate in regimes where the total detection rate is 20$\%$ to 30$\%$.
 
\bibliographystyle{naturemag}
\bibliography{refs_2.bib}

\newpage
\textbf{Supplementary Information} is available in the online version of the paper.

\textbf{Acknowledgments} 
This work was funded by NSERC, CIFAR, and Northrop-Grumman Aerospace Systems. We would like to acknowledge Alan Stummer's design and construction of several electronic devices which where essential to this experiment.

\textbf{Author Contributions}
All authors contributed equally to the results,
interpretation and presentation of this Article.

\textbf{Author Information}
Reprints and permissions information is available at
www.nature.com/reprints. The authors declare no competing financial interests.
Readers are welcome to comment on the online version of the paper. Correspondence
should be addressed to M. H. (mhallaji@physics.utoronto.ca)

\newpage
\appendix
\emph{Experimental parameters.}
Table \ref{tab:parameter} shows the mean incident photon number, post-selection parameter, overall detection efficiency, total number of measurement trials (clicks and no-clicks), and measured post-selection probability for each data point used in Fig\ref{yes_no}.

\begin{table}[h]
\begin{center}
\begin{tabular}{ | l | l | l | l | l | l |}
\hline Data Point & $\bar{n}$ & $\delta$ & $\eta$ & N$_{\mathrm{tot}}$ (approx) & P$_{\mathrm{click}}$\\
\hline 1 & 95 & 0.10 & 0.2 & 42,111,000 & 31\% \\
\hline 2 & 45 & 0.14 & 0.2 & 104,111,000 & 23\% \\
\hline 3 & 20 & 0.22 & 0.2 & 102,380,000 & 25\% \\
\hline 4 & 10 & 0.32 & 0.2 & 204,112,000 & 23\% \\
\hline 5 & 40 & 1 & 0.03 & 374,443,000 & 20\%\\
\hline
\end{tabular}
\caption{Experimental parameters used in the WVA experiment.}
\label{tab:parameter}
\end{center}
\end{table}

For each data point, the post-selection parameter, detector efficiency and mean photon number in signal were set so that the expected post-selection probability was about 19\%, not including background photons.     
Table \ref{tab:parameter} shows the measured probability of ``click'' for each data point.
As you can see, except for data points number 1 and 5, we typically observed click rates of approximately 25\%, due to the additional 6\% background photons.
For data point number 5, we used an ND filter to reduce the overall collection efficiency of the detector, and as a result, the background counts dropped to around 1-2\%.
For data point 1 we measured the expected 6\% from background photons (with the signal beam off) and 20\% from the signal pulses (with the probe blocked).
Yet the total measured rate in the presence of the two beams was about 30\%. 
The reason for this is a bit subtle.
We believe that the main source of the background photons is the scattered probe and coupling photons; if the probe sees a higher OD, it will scatter more photons and as a result the background rate will increase.
For data point 1, we sent signal pulses with 95 photons on average. 
The ac-Stark shift due to 95 photons shifts the probe about 100KHz outside the transparency window, and as a result the probe sees a higher OD and therefore scatters more photons.
The fact that these added background photons only exist in the presence of the signal pulses makes their full characterization difficult.

\begin{figure}[h]
\centering
\includegraphics[width = 0.6\columnwidth]{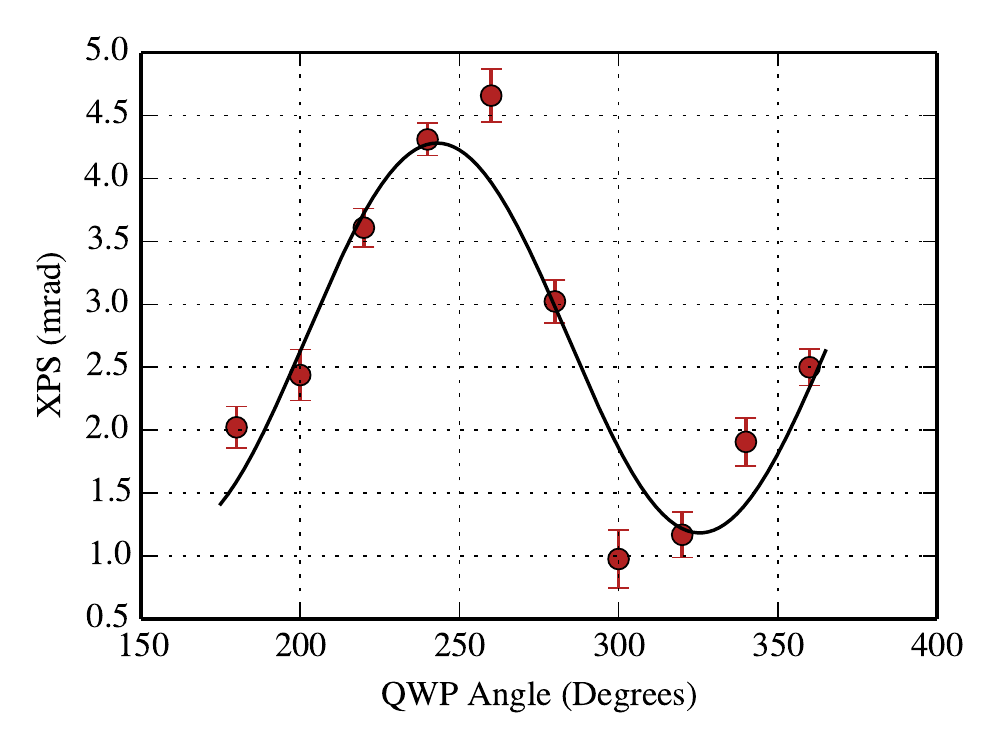}
\caption{  
  \label{sup_3}}
\end{figure}

\begin{figure}[h]
\centering
\includegraphics[width = 0.6\columnwidth]{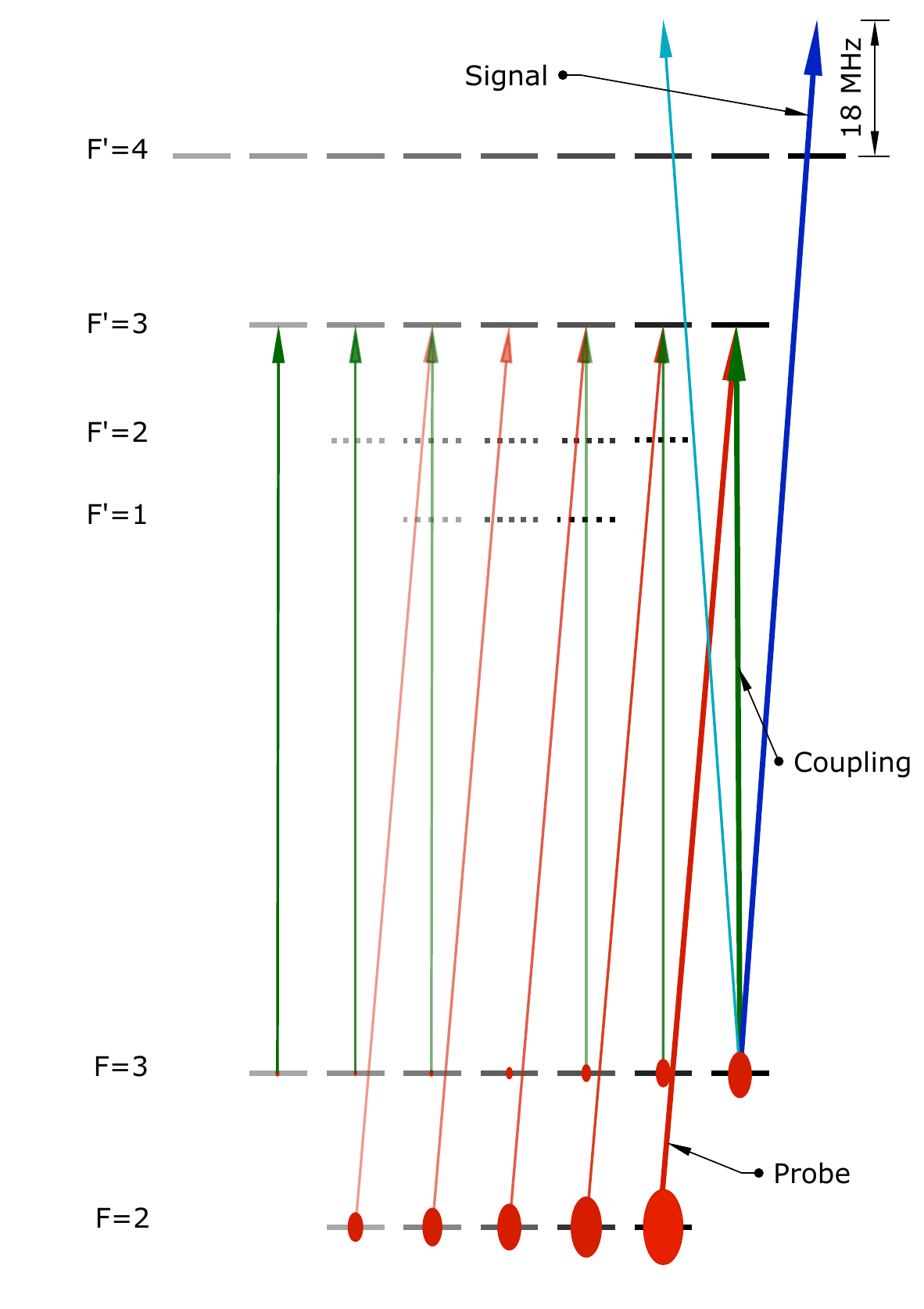}
\caption{ 
  \label{sup_2}}
\end{figure}

\emph{Level Scheme}.
The WVA experiment relies on a difference between the interaction strength for the probe with each of the signal ``paths'' (polarizations, in our implementation).
Therefore, we chose a level structure in which the probe interacts more strongly with the $\sigma^+$-polarized signal than with the $\sigma^-$-polarized signal.
Figure \ref{sup_2} shows the level scheme we used.
The reason for this polarization dependence can be understood as follows:
the magnitude of the XPS is proportional to $\Omega_s^2 \Omega_{pr}^2 / \Omega_{c}^2$ (in the low OD limit) where the $\Omega_i$'s are the Rabi frequencies of the signal, probe and coupling beams.
The Clebsch-Gordon (CG) coefficient for $\sigma^+$-polarized probe is largest for the $F=2,m_F = 2 \to F'=3,m_{F'}=3$ transition.
Likewise, the CG coefficient for the $\pi$-polarized coupling beam is larger for the $F=3,m_F = 3 \to F'=3,m_{F'}=3$ transition than the $F=3,m_F = 2 \to F'=3,m_{F'}=2$ transition.
Therefore, the $\Lambda$-system created between the $F=2,m_F = 2$ and $F=3,m_F = 3$ ground states is the dominant $\Lambda$-system.
One can easily show that for equal population distribution among the $F=2$ ground states, one should expect the XPS for the $\sigma^+$-polarized component of the signal to be around 3.6 times larger than the XPS for the $\sigma^-$-polarized component.
If one initialized all the population in the $F=2,m_F=2$ ground state (via optical pumping), the ratio between the two XPS's would be 28.
For this experiment, although we do not have a dedicated optical pumping stage in the measurement cycle, the probe beam is expected to pump the atomic population towards Zeeman levels with larger $m_F$'s, which is the reason for the observed ratio of 6.5 for $\phi_+ / \phi_-$ ( $\phi_+ = 10.0 \pm 0.6 \mu$rad and $\phi_- = 1.2 \pm 0.6 \mu$rad), as opposed to 3.6.
Fig.\ref{sup_3} shows the measured XPS versus the angle of the quarter-wave plate that controls the initial polarization of the signal.
The dependence of the XPS on the signal polarization can be clearly seen.

\begin{figure}[h]
\centering
\includegraphics[width = 0.6\columnwidth]{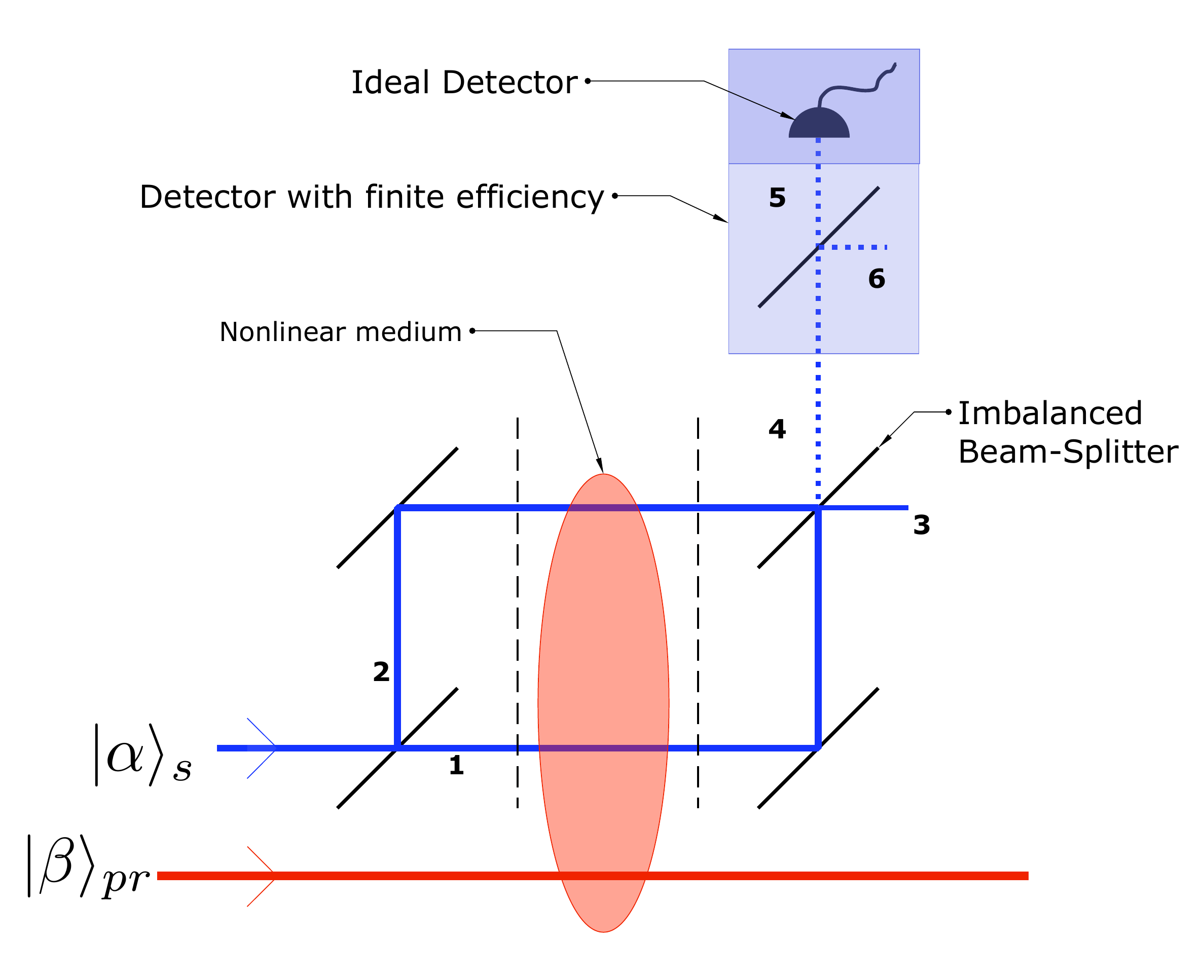}
\caption{ 
  \label{sup_1}}
\end{figure}

\emph{Expected XPM written on the probe -- full quantum mechanical calculation}.
\noindent
Here we calculate the expected nonlinear phase shift written on the probe by the signal in the geometry shown in Fig. \ref{sup_1}.
For now, we assume 100\% efficiency for the single-photon detector.
For a coherent state signal $| \alpha \rangle _s$ and a coherent state probe $|\beta \rangle _{pr}$, the combined state of the signal and the probe before the interaction is
\begin{equation}
  |\psi\rangle_{\mathrm{before}} = |\alpha / \sqrt{2}\rangle_1 |\alpha / \sqrt{2}\rangle_2 |\beta \rangle _{pr},
\end{equation}
\noindent
where $1$ and $2$ denote the two ``paths'' ($\sigma_+$ and $\sigma_-$) of the interferometer, between which the amplitude $\alpha$ is split evenly.
The interaction can be modeled via unitary propagator as $U = \exp \{i(\phi_1 n_1 n_{pr} + \phi_2 n_2 n_{pr})\}$, where $n_1$,$n_2$ and $n_{pr}$ are the number operators for the fields in paths $1$,$2$ and probe, respectively.
$\phi_1$ and $\phi_2$ denote the interaction strengths in each arm.
Here we assume $\phi_1 > \phi_2$.
After the interaction the combined state can be written via 

\begin{equation}
  |\psi\rangle_{i} = \sum_{n}^{} e^{-|\beta|^2 /2} \frac{\beta^n}{\sqrt{n!}} \,|\frac{\alpha}{\sqrt{2}} e^{in\phi_1}\rangle_1 |\frac{\alpha}{\sqrt{2}} e^{in\phi_2}\rangle_2 |n \rangle _{pr}.
\end{equation}
\noindent
The imbalanced beam-splitter can be modeled as 
\begin{equation}
  \label{mtrx}
  \begin{pmatrix}
  a_3\\
  a_4
  \end{pmatrix}
  = 
  \begin{pmatrix}
  \cos \theta & -\sin \theta \\
  \sin \theta & \cos \theta
  \end{pmatrix}
  \begin{pmatrix}
  a_1\\
  a_2
  \end{pmatrix}.
\end{equation}
\noindent
Therefore, the combined state after the interferometer can be written as:
\begin{equation}
\begin{split}
  \label{after_int}
  |\psi\rangle_{f} = \sum_{n}^{} e^{-|\beta|^2 /2} \frac{\beta^n}{\sqrt{n!}} & \left|\frac{\alpha e^{in\bar{\phi}}}{\sqrt{2}} (e^{in\Delta \phi /2} \cos \theta - e^{-in\Delta \phi /2}\sin \theta) \right\rangle_3 \\
  & \left|\frac{\alpha e^{in\bar{\phi}}}{\sqrt{2}} (e^{in\Delta \phi /2} \sin \theta + e^{-in\Delta \phi /2}\cos \theta) \right\rangle_4 \left|n \right\rangle _{pr}.
\end{split} 
\end{equation}
\noindent
where $\bar{\phi} = \frac{\phi_1 + \phi_2}{2}$ and $\Delta \phi = \phi_1 - \phi_2$.
We take the limit where the overlap between the initial and final states is very close to 1, and port 4 in Fig.\ref{sup_1} is nearly dark.
This implies that in eq.\ref{mtrx}, $\theta \approx -\pi/4$.
For $\theta + \pi/4 \ll 1$, we expand $\cos \theta$ and $\sin \theta$ as $\cos \theta = \frac{1- \delta}{\sqrt{2}}$ and $\sin \theta =- \frac{1+ \delta}{\sqrt{2}}$.
Therefore, the state in eq.\ref{after_int} becomes
\begin{equation}
\begin{split}
\label{after_approx:1}
|\psi \rangle _{f} = \sum_{n}^{} e^{-|\beta|^2 /2} \frac{\beta^n}{\sqrt{n!}}& \left| \frac{\alpha e^{in\bar{\phi} } }{2} ( e^{ in\Delta \phi /2 }(1-\delta) + e^{ -in\Delta \phi /2 }(1+\delta) ) \right \rangle_3 \\
& \left|\frac{\alpha e^{in\bar{\phi} } }{2} ( e^{ in\Delta \phi /2 }(1+\delta) - e^{ -in\Delta \phi /2 }(1-\delta) )\right \rangle_4 |n \rangle _{pr}.
\end{split}
\end{equation} 
If $\bar{n}_{pr}$ is sufficiently small so that terms with $n_{pr} \geq 1/ \Delta \phi$ can be neglected, we can expand $e^{\frac{\pm in\Delta \phi}{2}}$ as $1 \pm \frac{in\Delta \phi}{2}$.
Applying this expansion in eq.\ref{after_approx:1} and keeping terms to the first order in $\Delta \phi$ and $\delta$ we have
\begin{equation}
\label{after_approx:2}
|\psi \rangle _{f} = \sum_{n}^{} e^{-|\beta|^2 /2} \frac{\beta^n}{\sqrt{n!}} | \alpha e^{in\bar{\phi}} \rangle_3 | \alpha \delta e^{in\bar{\phi}} ( 1 + \frac{in\Delta \phi}{2 \delta} ) \rangle_4 |n \rangle _{pr}.
\end{equation}
Assuming $n\Delta \phi \ll \delta$, we can write $1 + \frac{in\Delta \phi}{2 \delta}$ as $e^{\frac{in\Delta \phi}{2\delta}}$.
So we have
\begin{equation}
\label{after_approx:3}
|\psi \rangle _{f} = \sum_{n}^{} e^{-|\beta|^2 /2} \frac{\beta^n}{\sqrt{n!}} | \alpha e^{in\bar{\phi}} \rangle_3 | \alpha \delta e^{in\bar{\phi}} e^{\frac{in\Delta \phi}{2\delta}} \rangle_4 |n \rangle _{pr}.
\end{equation}
For $|\alpha|^2 \delta^2 \ll 1$, one can write the state in arm $4$, the dark port, as $| \alpha \delta e^{in\bar{\phi}} e^{in\Delta \phi/2\delta} \rangle_4 = |0\rangle _4 + \alpha \delta e^{in (\bar{\phi} +\Delta \phi/2\delta )}  |1\rangle_4 $.
To calculate what happens when the single-photon detector does not fire, we project onto $|0\rangle _4$, and find that the state of the signal and probe becomes
\begin{equation}
\label{reduced:1}
|\psi \rangle _{no-click} = _4\langle 0 |\psi \rangle _{f} = \sum_{n}^{} e^{-|\beta|^2 /2} \frac{\beta^n}{\sqrt{n!}} | \alpha e^{in\bar{\phi}} \rangle_3 |n \rangle _{pr}
\end{equation}
and after expanding the coherent state in mode $3$ we have
\begin{equation}
\label{reduced:2}
|\psi \rangle _{no-click} = \sum_{n,m}^{} e^{-|\beta|^2 /2} \frac{\beta^n}{\sqrt{n!}} e^{-|\alpha|^2 /2} \frac{\alpha^m}{\sqrt{m!}} e^{inm\bar{\phi}} | m \rangle_3 |n \rangle _{pr},
\end{equation}
which can be written as
\begin{equation}
\label{reduced:3}
|\psi \rangle _{no-click} = \sum_{m}^{} e^{-|\alpha|^2 /2} \frac{\alpha^m}{\sqrt{m!}} | m \rangle_3 |\beta  e^{im\bar{\phi}} \rangle _{pr}.
\end{equation}
This result shows that, for each $m$-photon component of the signal state, the nonlinear phase shift of the probe is $m\bar{\phi}$, as expected.
Therefore the average phase shift on the probe will be
\begin{equation}
\label{avgphi:1}
\bar{\phi}_{no-click} = \sum_{m}^{} P(m) m\bar{\phi} = |\alpha|^2 \bar{\phi}
\end{equation}

On the other hand, when the detector in the dark port fires, we project onto $|1\rangle _4$ and find
\begin{equation}
\label{reduced:4}
\begin{split}
|\psi \rangle _{click} &= \sum_{n}^{} e^{-|\beta|^2 /2} \frac{\beta^n}{\sqrt{n!}} e^{in(\bar{\phi} + \frac{\Delta \phi}{2 \delta} )} | \alpha e^{in\bar{\phi}} \rangle_3 |n \rangle _{pr} \\
&= \sum_{m}^{} e^{-|\alpha|^2 /2} \frac{\alpha^m}{\sqrt{m!}} | m \rangle_3 | \beta e^{im\bar{\phi} + i(\bar{\phi} + \frac{\Delta \phi}{2 \delta}) } \rangle _{pr}. 
\end{split}
\end{equation}
Therefore, the phase shift on the probe can be shown to be
\begin{equation}
\label{avgphi:2}
\bar{\phi}_{click} = \sum_{m}^{} P(m) ( \bar{\phi}(m+1) + \frac{\Delta \phi}{2 \delta} ) = (|\alpha|^2 +1)\bar{\phi} + \frac{\Delta \phi}{2 \delta}
\end{equation}

It is worth noting that this result is obtained under the condition that $\bar{n}_{probe} \Delta \phi \ll \delta $.
That is to say, the post-selection parameter must be larger than the phase-shift written on the signal due to the probe.
This condition dictates that the probability of a successful post-selection should be dominated by the imbalance in the interferometer introduced as $\delta$ and not the false positives due to the nonlinear phase shift $n \Delta \phi$ inside the interferometer (interaction back-action).
It is easy to show that when $\bar{n}_{probe} \Delta \phi \gg \delta $, no anomalous result should be expected.

\emph{Weak value of photon numbers -- weak value approach}.
\noindent
Now we calculate the weak value of the photon number in each arm of the interferometer versus the post selection parameter $\delta = -(\cos \theta + \sin \theta ) / \sqrt{2}$, where $\theta$ is introduced in eq.\ref{mtrx}.
In fig.\ref{sup_1}, the single-photon detector with efficiency $\eta$ placed in the dark port (mode 4) is modeled as a beam-splitter which transmits a fraction $\eta$ of the incoming light to the ``detected'' mode (mode 5 in figure \ref{sup_1}), followed by an ideal detector.
The remaining fraction ($1-\eta$) of the incoming light is reflected to the ``undetected'' mode (mode 6).
The initial state of the signal can be written in modes 3, 5 and 6 as
\begin{equation}
\label{ini}
|i\rangle = |\alpha\rangle_s = \Big|\frac{\alpha}{\sqrt{2}}(\cos \theta - \sin \theta)\Big\rangle_3 \Big|\frac{\alpha \sqrt{(1-\eta)} }{\sqrt{2}}(\cos \theta + \sin \theta)\Big\rangle_{6} \Big|\frac{\alpha \sqrt{\eta}}{\sqrt{2}}(\cos \theta + \sin \theta)\Big\rangle_{5}. 
\end{equation} 
After the interaction, the state in the ``detected'' mode is projected onto $|1\rangle_{5}$.
The resulting state after the post-selection is
\begin{equation}
\label{final}
|f\rangle = |\frac{\alpha}{\sqrt{2}}(\cos \theta - \sin \theta)\rangle_3 |\frac{\alpha \sqrt{(1-\eta)} }{\sqrt{2}}(\cos \theta + \sin \theta)\rangle_{6} |1\rangle_{5}. 
\end{equation} 
The weak values of the photon number in the two arms of the interferometer are given by
\begin{equation}
\label{wkvl}
\langle n_{1(2)} \rangle _{wk} = \frac{\langle f | a^{\dagger}_{1(2)}a_{1(2)} | i \rangle }{\langle f | i \rangle},
\end{equation}
where $a^{\dagger}_{1(2)}$ and $a_{1(2)}$ are the creation and annihilation operators for modes 1(2).
Using the transformation matrix in eq.\ref{mtrx} and the fact that the beam-splitter (which models the imperfect detector) in mode 4 has transmissivity $t^2 = \eta$ and reflectivity $r^2 = 1-\eta$, the photon number operator in mode 1, $\hat{n}_1 = a^{\dagger}_1 a_1$, can be written in terms of those of modes 3, 5, and 6 as
\begin{equation}
\label{n_1}
\begin{split}
\hat{a}^{\dagger}_1\hat{a}_1  =& \cos^2 \theta \hat{a}^{\dagger}_3\hat{a}_3 + r \sin \theta \cos\theta \hat{a}^{\dagger}_3\hat{a}_6 + t \sin\theta\cos\theta\hat{a}^{\dagger}_3\hat{a}_5\\
&r \sin\theta\cos\theta \hat{a}^{\dagger}_6\hat{a}_3 + r^2 \sin^2\theta \hat{a}^{\dagger}_6\hat{a}_6 + rt \sin^2\theta\hat{a}^{\dagger}_6\hat{a}_5\\
&t \sin\theta\cos\theta\hat{a}^{\dagger}_5\hat{a}_3 + rt \sin^2\theta\hat{a}^{\dagger}_5\hat{a}_6 +t^2\sin^2\theta\hat{a}^{\dagger}_5\hat{a}_5.    
\end{split}
\end{equation}
The photon number operator in mode 2 can be similarly calculated. 
Using the initial and final states introduced above, one can calculate the weak values to be
\begin{equation}
\begin{split}
\langle n_{1} \rangle _{wk} &= \frac{|\alpha|^2}{2} + \frac{\sin\theta}{\cos\theta+\sin\theta} - \frac{\eta|\alpha|^2}{2}[ \sin \theta (\sin \theta + \cos \theta)] \\
\langle n_{2} \rangle _{wk} &= \frac{|\alpha|^2}{2} + \frac{\cos\theta}{\cos\theta+\sin\theta} - \frac{\eta|\alpha|^2}{2}[ \cos \theta (\sin \theta + \cos \theta)].
\end{split} 
\end{equation} 
In the limit of small overlap between the initial and final state ($\theta \approx - \pi/4$ and $\eta |\alpha|^2 \ll 1$), we have $\cos \theta = \frac{1- \delta}{\sqrt{2}}$ and $\sin \theta =- \frac{1+ \delta}{\sqrt{2}}$.
Therefore in this limit the weak values become 
\begin{equation}
\begin{split}
\langle n_{1} \rangle _{wk} &= \frac{|\alpha|^2}{2} + \frac{1}{2} + \frac{1}{2 \delta} \\
\langle n_{2} \rangle _{wk} &= \frac{|\alpha|^2}{2} + \frac{1}{2} - \frac{1}{2 \delta}.
\end{split} 
\end{equation}
Now if the single-photon phase shift in arms 1 and 2 are $\phi_1$ and $\phi_2$, the expected phase shift on the probe will be
\begin{equation}
\begin{split}
\bar{\phi}_{probe} &= \langle n_{1} \rangle _{wk}\phi_1 + \langle n_{2} \rangle _{wk} \phi_2 \\
&= (|\alpha|^2 + 1) (\frac{\phi_1 + \phi_2}{2}) + \frac{\phi_1 - \phi_2}{2 \delta}\\
&=(|\alpha|^2+1)\bar{\phi} + \frac{\Delta \phi}{2 \delta}
\end{split} 
\end{equation}
which recovers the result of eq.\ref{avgphi:2}.

\textbf{Extended data}

\emph{Figure \ref{sup_3}}.
XPS versus signal polarization. 
A quarter-wave plate is used to change the polarization of the signal beam before the interaction.
The maximum and minimum measured phase shifts correspond to circular polarizations $\sigma^+$ and $\sigma^-$, respectively.
The signal pulses used for this measurement contained around 600 photons.

\emph{Figure \ref{sup_2}}.
The full level structure used in the experiment.

\emph{Figure \ref{sup_1}}.
Schematic of interferometer.

\newpage

\end{document}